\documentclass[prc,twocolumn,showpacs,showkeys,nofootinbib,superscriptaddress]{revtex4-1}
\usepackage{graphicx}
\usepackage{dcolumn}
\usepackage{epsfig}
\usepackage{bm}

\usepackage{amsfonts}
\usepackage{amsmath}
\usepackage{amssymb}
\usepackage[english]{babel}
\usepackage{graphicx}
\usepackage{epsfig}
\usepackage{bm}
\usepackage{verbatim} 
\usepackage[utf8]{inputenc}
\usepackage{booktabs}
\usepackage{multirow}
\usepackage[dvipsnames,table,xcdraw]{xcolor}
\usepackage[colorlinks=true,urlcolor=blue,citecolor=blue]{hyperref}
\usepackage{float}
\usepackage{blindtext}

\begin{document}
	
	\title{Radiative and exchange corrections for two-neutrino double-beta decay}
	
	\author{O. Ni\c{t}escu}
	\email{ovidiu.nitescu@nipne.ro}
	\affiliation{Faculty of Mathematics, Physics and Informatics, Comenius University in Bratislava, 842 48 Bratislava, Slovakia}
	\affiliation{ International Centre for Advanced Training and Research in Physics, P.O. Box MG12, 077125 M\u{a}gurele, Romania}
	\affiliation{“Horia Hulubei” National Institute of Physics and Nuclear Engineering, 30 Reactorului, POB MG-6, RO-077125 Bucharest-M\u{a}gurele, Romania}
	
	\author{F. \v{S}imkovic}
	\email{fedor.simkovic@fmph.uniba.sk}
	\affiliation{Faculty of Mathematics, Physics and Informatics, Comenius University in Bratislava, 842 48 Bratislava, Slovakia}
	\affiliation{Institute of Experimental and Applied Physics, Czech Technical University in Prague,
		110 00 Prague, Czech Republic}
	
	\begin{abstract}
		We investigate the impact of radiative and atomic exchange corrections in the two-neutrino double-beta ($2\nu\beta\beta$)-decay of $^{100}$Mo. In the calculation of the exchange correction, the electron wave functions are obtained from a modified Dirac-Hartree-Fock-Slater self-consistent framework that ensures orthogonality between continuum and bound states. The atomic exchange correction causes a steep increase in the low-energy region of the single-electron spectrum, consistent with previous studies on $\beta$-decay, while the radiative correction primarily accounts for a 5\% increase in the decay rate of $^{100}$Mo. When combined, the radiative and exchange effects cause a leftward shift of approximately 10 keV in the maximum of the summed electron spectrum. This shift may impact current constraints on parameters governing potential new physics scenarios in $2\nu\beta\beta$-decay. The exchange and radiative corrections are introduced on top of our previous description of $2\nu\beta\beta$-decay, where we used a Taylor expansion for the lepton energy parameters within the nuclear matrix elements denominators. This approach results in multiple components for each observable, controlled by the measurable $\xi_{31}$ and $\xi_{51}$ parameters. We explore the effects of different $\xi_{31}$ and $\xi_{51}$ values, including their experimental measurements, on the total corrected spectra. These refined theoretical predictions can serve as precise inputs for double-beta decay experiments investigating standard and new physics scenarios within $2\nu\beta\beta$-decay.
	\end{abstract}
	
	\maketitle

	\section{Introduction}
	
	Two-neutrino double beta decay ($2\nu\beta\beta$-decay) is a process in which two neutrons in a nucleus $(A, Z_i)$, where $A$ and $Z_i$ are the mass and atomic numbers, respectively, decay simultaneously into two protons, emitting two electrons and two electron antineutrinos \cite{Goeppert-Mayer-PR1935},
	\begin{equation}
		(A, Z_i) \rightarrow (A, Z_i+2) + e^- + e^- + \overline{\nu}_e  + \overline{\nu}_e.
	\end{equation}
	This decay is allowed within the framework of the Standard Model (SM) and it is the rarest observed nuclear decay process with half-lives exceeding $10^{18}$ years \cite{Barabash-U2020}.  In contrast, the unobserved neutrinoless double beta decay ($0\nu\beta\beta$-decay), which involves the same initial and final states but does not emit any neutrinos \cite{Furry-PR1939}, would indicate a violation of lepton number. The discovery of $0\nu\beta\beta$-decay would establish the Majorana nature of neutrinos \cite{Majorana-INC1937, Schechter-PRD1982} and provide crucial insights into neutrino masses, their hierarchy, and charge-parity violation in the lepton sector \cite{Schechter-PRD1982,Sujkowski-PRC2004,Pascoli-NPB2006,Bilenky-IJMPA2015,Vergados-IJMPE2016,Girardi-NPB2016,Simkovic-PU2021}.
	
	One of the theoretical challenges in $2\nu\beta\beta$-decay and $0\nu\beta\beta$-decay is the calculation of the nuclear matrix elements (NMEs), a long-standing problem in this field. The difficulty arises because the nuclei involved in $\beta\beta$-decay are typically open-shell medium and heavy nuclei with complex structures. Moreover, for the two-neutrino mode, a complete set of $1^+$ intermediate nucleus states must be described, while for the neutrinoless mode, contributions from all multipolarities are required. The current NMEs predictions rely on various nuclear structure models, such as the proton-neutron quasiparticle random phase approximation (pn-QRPA) and its variants \cite{Vogel-PRL1986,Civitarese-PLB1987,Aunola-NPA1996a,Aunola-NPA1996b,Suhonen-PR1998,Stoica-NPA2001,Stoica-TEPJA2003,Simkovic-NPA2004,Alvarez-PRC2004,Rodin-NPA2006,Fang-PRC2011,Suhonen-JPG2012,Simkovic-PRC2013,Mustonen-PRC2013,Hyvarien-PRC2015,Pirinen-PRC2015,Simkovic-PRC2018,Fang-PRC2018}, the nuclear shell model (NSM) \cite{Caurier-PRL1996,Caurier-RMP2005,Horoi-PRC2007,Menendez-NPA2009,Caurier-PLB2012,Neacsu-PRC2015,Brown-PRC2015,Horoi-PRC2016,Kostensalo-PLB2020,Coraggio-PRC2020,Patel-NPA2024,Fang-PRC2024}, the interacting boson model (IBM) \cite{Barea-PRC2013,Barea-PRC2015,Nomura-PRC2022,Nomura-PRC2024}, the projected Hartree-Fock-Bogoliubov (PHFB) method \cite{Rath-PRC2013,Rath-FP2019}, effective theory (ET) \cite{CoelloPerez-PRC2018,Brase-PRC2022}, and others \cite{Kotila-JPG2010,Rodriguez-PLB2013,Menendez-PRC2014,Popara-PRC2022}. We also note the recent efforts toward the \textit{ab initio} NMEs calculations starting from realistic nuclear forces \cite{Yao-PRL2020,Belley-PRL2021,Yoa-PRC2021,Novario-PRL2021,Coraggio-PRC2024} and, at extreme, the simplifications introduced by the phenomenological models \cite{Ren-PRC2014,Rajan-IJP2018,Pritychenko-NPA2023,Nitescu-arXiv2024}.

	In addition to NMEs, the precision of observables such as the single and summed energy distributions and angular correlation between the emitted electrons is crucial for understanding various hypothesis in $2\nu\beta\beta$-decay \cite{Simkovic-JPG2001,Domin-NPA2005} and for unraveling the underlying mechanism driving $0\nu\beta\beta$-decay \cite{Neacsu-AHEP2016,Horoi-PRD2016,Simkovic-PRC2018,Graf-PRD2022,Scholer-JHEP2023}. Additionally, the current experimental constraints on various strength parameters associated with the beyond the Standard Model (BSM) scenarios are obtained by analyzing the shape of the summed electron energy distribution of $2\nu\beta\beta$-decay \cite{Bossio-JPG2024}. Moreover, the SM predictions for the $2\nu\beta\beta$-decay also play an important role in the experimental searches for weakly interacting massive particles (WIMPs) and coherent elastic neutrino-nucleus scattering (CE$\nu$NS). In particular, in liquid Xenon experiments, the $2\nu\beta\beta$-decay of $^{136}$Xe represents an inevitable source of background. Therefore, precise theoretical predictions for $2\nu\beta\beta$-decay are necessary for several upcoming experiments aiming to detect WIMPs and CE$\nu$NS \cite{XENONnT-JCAP2020,LUX-ZEPLIN-PRD2020,DARWIN-JCAP2016,DARWIN-JPG2022}. 
	
	In this paper, we advance the precision of theoretical predictions for the observables of $2\nu\beta\beta$-decay by considering the radiative and atomic exchange effects for the emitted electrons. Our primary focus is on the $2\nu\beta\beta$-decay of $^{100}$Mo, but similar results are expected for other nuclei undergoing $2\nu\beta\beta$-decay. The corrections are introduced on top of our previously improved description of the $2\nu\beta\beta$-decay \cite{Simkovic-PRC2018, Nitescu-U2021}, where we employed a Taylor expansion on the parameters encompassing the lepton energies within the denominators of the NMEs. This formalism enables decomposition of $2\nu\beta\beta$-decay observables into partial components governed by the parameters $\xi_{31}$ and $\xi_{51}$. The connection of the Taylor expansion formalism with the the single state dominance (SSD) and the higher state dominance (HSD) hypotheses is presented.   
	
	For the radiative correction, we account the exchange of a virtual photon or the emission of a real photon during $2\nu\beta\beta$-decay. For the atomic exchange correction, we account for the possibility that the electron emitted from the decay could be exchanged with an atomic bound electron. Simultaneously, the bound electron undergoes a transition to a continuum state of the final atom. For the description of the bound and continuum states, required in the calculation of the atomic exchange correction, we employ the Dirac-Hartree-Fock-Slater (DHFS) self-consistent framework. Our findings reveal a roughly 5\% increase in the $2\nu\beta\beta$-decay rate of $^{100}$Mo, mostly due to radiative correction. Additionally, we observe (i) a steep increase in low-energy single-electron events due to the exchange correction, aligning with previous $\beta$-decay studies \cite{Harston-PRA1992,Pyper-PRSLA1998,Nitescu-PRC2023,Nitescu-PRC2024}, and (ii) a 10 keV leftward shift in the maximum of the summed electron spectrum as a combined result of both corrections. Finally, we present the total corrected single and summed electron spectra for $2\nu\beta\beta$-decay of $^{100}$Mo evaluated for three parameter sets of $\xi_{31}^{2\nu}$ and $\xi_{51}^{2\nu}$ under the HSD hypothesis, the SSD hypothesis, and based on experimental measurements.

	\section{$2\nu\beta\beta$-decay rate within different hypothesis}

	\subsection{Taylor expansion formalism}

	In our previous studies \cite{Simkovic-PRC2018,Nitescu-U2021}, we shown that the lepton energies within the denominators of the NMEs, which is usually neglected, can be accounted for in the $2\nu\beta\beta$-decay through the application of a Taylor expansion on the parameters $\varepsilon_{K,L}$. By restricting our analysis to the fourth power of $\varepsilon_{K,L}$, the $2\nu\beta\beta$-decay rate can be written as \cite{Simkovic-PRC2018,Nitescu-U2021},
	
	\begin{eqnarray}
		\Gamma^{2\nu} = \Gamma_0^{2\nu} + \Gamma_2^{2\nu}  + \Gamma_{22}^{2\nu} + \Gamma_4^{2\nu}, 
	\end{eqnarray}
	where, the leading $\Gamma^{2\nu}_0$, next to leading $\Gamma^{2\nu}_2$ and next-to-next to leading $\Gamma^{2\nu}_{22}$ and $\Gamma^{2\nu}_4$ orders of the Taylor expansion are given by
	
	\begin{align}
		\begin{aligned}
			\frac{\Gamma_0^{2\nu}}{\ln{(2)}}
			&= \left(g^{\rm eff}_A\right)^4 {\cal M}_{0} G^{2\nu}_{0},~
			\frac{\Gamma_2^{2\nu}}{\ln{(2)}}
			= \left(g^{\rm eff}_A\right)^4 {\cal M}_{2} G^{2\nu}_{2}, \\ \frac{\Gamma_{22}^{2\nu}}{\ln{(2)}}
			&= \left(g^{\rm eff}_A\right)^4 {\cal M}_{22} G^{2\nu}_{22},
			\frac{\Gamma_4^{2\nu}}{\ln{(2)}}
			= \left(g^{\rm eff}_A\right)^4 {\cal M}_{4} G^{2\nu}_{4},
		\end{aligned}
	\end{align}
	respectively. Here $g^{\rm eff}_A$ is the effective axial-vector coupling constant which is still an open question in nuclear weak interaction processes \cite{Suhonen-FP2017}. The quantities ${\cal M}_{N}$, with $N=\{0,2,22,4\}$, defined in \cite{Simkovic-PRC2018}, are combinations of the standard GT component of the NME of $2\nu\beta\beta$-decay,
	
	\begin{equation}
		M^{2\nu}_{GT-1}\equiv M^{2\nu}_{GT}=\sum_{n}M_n\frac{m_e}{E_n-(E_i+E_f)/2},
	\end{equation}
	and the new NMEs arising from the Taylor expansion,
	
	\begin{align}
		\begin{aligned}
			M^{2\nu}_{GT-3} &= \sum_n M_n\frac{4~ m_e^3}{\left[E_n - (E_i+E_f)/2\right]^3},\\ 
			M^{2\nu}_{GT-5} &= \sum_n M_n\frac{16~ m_e^5}{\left[E_n - (E_i+E_f)/2\right]^5}.
		\end{aligned}
	\end{align}
	Here, $M_n$ is dependent on the n$^{\rm th}$ $1^+$ state of the intermediate nucleus, with energy $E_n$, which contributes to the summation,
	
	\begin{equation}
		M_n =
		\langle 0^+_f\|\sum_{j}\tau^+_j\sigma_j\| 1^+_n\rangle\langle 1^+_n\|\sum_{k}\tau^+_k\sigma_k\| 0^+_i\rangle.
	\end{equation}
	Here, $|0^+_i\rangle$ ($|0^+_f\rangle$) is the ground state of the initial (final) even-even nucleus with energy $E_i$ ($E_f$), and the summations run over all nucleons. The operator $\sigma_{j,k}$ is the nucleon spin operator, and $\tau^+_{j,k}$ represents the isospin-ladder operator transforming a neutron into a proton. We adopt the same assumptions as in \cite{Simkovic-PRC2018}, where the Fermi (F) component of the NME for $2\nu\beta\beta$ decay is neglected. However, it should be noted that the expressions for ${\cal M}_{N}$, with $N=\{0,2,22,4\}$, which take into account the F component, are presented in \cite{Nitescu-U2021}.
	
	Taking advantage of the analytical integration over the antineutrino energy preformed in \cite{Nitescu-U2021}, the PSFs, in the Taylor expansion formalism, are given by,
	
	\begin{align}
		\label{eq:PhaseSpaceFactorsWithCorrections}
		\begin{aligned}
			G_N^{2\nu}&=\frac{(G_{F}\left|V_{ud}\right|)^4}{8\pi^7m_e^2\ln{2}}\int_{m_e}^{E_i-E_f-m_e}\int_{m_e}^{E_i-E_f-E_{e_1}}  \\ 
			&\times p_{e_1}E_{e_1}\left[1+\eta^T(E_{e_1})\right]R(E_{e_1},E_i-E_f-m_e)\\
			&\times p_{e_2}E_{e_2}\left[1+\eta^T(E_{e_2})\right]R(E_{e_2},E_i-E_f-E_{e_1}) \\
			&\times  F_{ss}(E_{e_1})F_{ss}(E_{e_2}) \mathcal{I}_{N} dE_{e_2}dE_{e_1}
		\end{aligned}
	\end{align}
	with $N=\{0,2,22,4\}$ and the dimensionless quantities $\mathcal{I}_N$ are the results of the analytical integrations over the antineutrino energy and are defined in the Appendix of \cite{Nitescu-U2021}. The energy $E_e$ is the total energy of the emitted electron with momentum $p_e$. $G_F$ is the Fermi coupling constant, $V_{ud}$ is the first element of the Cabibbo-Kobayashi-Maskawa (CKM) matrix and $m_e$ is the mass of the electron. The quantity $E_i-E_f$ represents the energy difference between the initial and final $0^+$ nuclear states, which can be determined by relating it to the $Q$-value of the $2\nu\beta\beta$-decay, given as $Q=E_i-E_f-2m_e$. For our study, we use $Q=3.0344$ MeV \cite{Qvalue-100Mo} for the $2\nu\beta\beta$-decay of $^{100}$Mo. Here, $E_{\nu_1}$ is the energy of one antineutrino and for the other $E_{\nu_2}=E_i-E_f-E_{e_1}-E_{e_2}-E_{\nu_1}$ from the energy conservation. The Fermi function, $F_{ss}$, is defined in \cite{Simkovic-PRC2018,Nitescu-U2021} and the functions $\left[1+\eta^T(E_{e})\right]$ and $R(E_{e},E_{e}^{\textrm{max}})$ account for the exchange and radiative corrections. These are discussed in Section~\ref{sec:ExRa}.

	By introducing ratios of the NMEs,
	
	\begin{equation}
		\xi_{31}=\frac{M_{GT-3}^{2\nu}}{M_{GT-1}^{2\nu}},\hspace{0.5cm}\xi_{51}=\frac{M_{GT-5}^{2\nu}}{M_{GT-1}^{2\nu}},
	\end{equation}
	one can write the $2\nu\beta\beta$-decay rate as a function of those ratios,
	\begin{widetext}
		\begin{align}
			\label{eq:TotalDecayRate}
			\begin{aligned}
				\left[T_{1/2}^{2\nu}(\xi_{31},\xi_{51})\right]^{-1}= \left(g^{\rm eff}_A\right)^4\left| M^{2\nu}_{GT-1}\right|^2 \left[G^{2\nu}_{0}+\xi_{31}G^{2\nu}_{2}+\frac{1}{3}\xi_{31}^2G^{2\nu}_{22}+\left(\frac{1}{3}\xi_{31}^2+\xi_{51}\right)G^{2\nu}_{4} \right].
			\end{aligned}
		\end{align}
	\end{widetext}

	The observables associated with $2\nu\beta\beta$-decay, such as the half-life and the single and summed electron distributions, are now dependent on the $\xi_{31}$ and $\xi_{51}$ parameters. From the theoretical point of view, one can predict those parameters by computing the NMEs within various nuclear models, e.g., pn-QRPA calculation with partial isospin restoration from \cite{Simkovic-PRC2018} or NSM and pn-QRPA calculations from \cite{KamLAND-Zen-PRL2019}. From an experimental perspective, the ratios $\xi_{31}$ and $\xi_{51}$ can be treated as free parameters that control the observables of $2\nu\beta\beta$-decay. The first limit on $\xi_{31}$ was set by the KamLAND-Zen collaboration by analyzing the $2\nu\beta\beta$-decay of $^{136}$Xe \cite{KamLAND-Zen-PRL2019}. Additionally, the $\xi_{31}$ and $\xi_{51}$ parameters have been recently explored by the CUPID-Mo experiment \cite{Augier-PRL2023-100Mo}. Combining the experimental limits with the theoretical predictions, conclusions can be drawn regarding the effective axial-vector coupling constant $g^{\rm eff}_A$ \cite{Simkovic-PRC2018}.
	
	\subsection{The SSD hypothesis}
	
	\begin{figure*}
		\centering
		\includegraphics[width=0.9\textwidth]{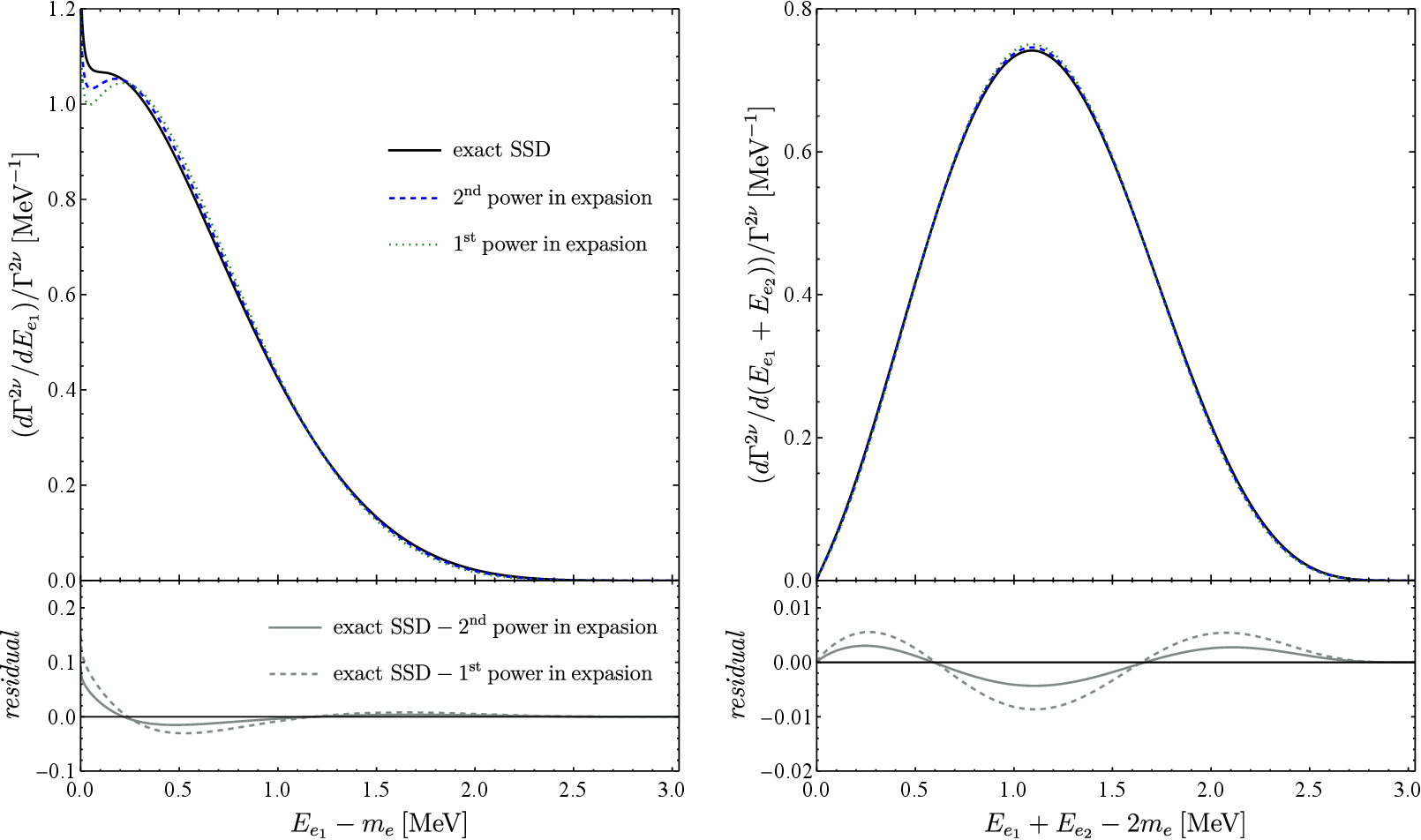}
		\caption{The single (left panel) and summed (right panel) electron spectra for $2\nu\beta\beta$-decay of $^{100}$Mo from: (solid) the exact SSD formalism, (dashed) Taylor expansion with contributions up to next-to-next to leading order, i.e., $\Gamma^{2\nu} = \Gamma_0^{2\nu} + \Gamma_2^{2\nu}  + \Gamma_{22}^{2\nu} + \Gamma_4^{2\nu}$, (dotted) Taylor expansion with contributions up to next to leading order, i.e., $\Gamma^{2\nu} = \Gamma_0^{2\nu} + \Gamma_2^{2\nu}$. All spectra are normalized to unity and they include radiative and exchange corrections. The lower panels display the residuals.  \label{fig:ExactSSDComparison}}
	\end{figure*}
	
	In the single-state dominance (SSD) hypothesis \cite{Abad-INCA1983,Abad-JPC1984}, it is assumed that the $2\nu\beta\beta$-decay is governed only by the transition through the first $1^+$ state of the intermediate nucleus with energy $E_1$. The inverse half-life for the $2\nu\beta\beta$-decay under the SSD hypothesis, for a $0^+\rightarrow0^+$ transition, can be written in the following form,
	
	\begin{equation}
		\label{eq:Half-lifeSSD}
		\left[T_{1/2}^{2\nu}\right]^{-1}=\left(g^{\rm eff}_A\right)^4\left|M^{2\nu}_{GT}(1)\right|^2 G^{2\nu}_{\rm{SSD}},
	\end{equation}
	where the square of the single state Gamow-Teller (GT) NME is given by
	
	\begin{equation}
		\left|M^{2\nu}_{GT}(1)\right|^2=\left|M_1^f(0^+)M_1^i(0^+)\right|^2,
	\end{equation}
	where we have used the notation from \cite{Simkovic-JPG2001,Domin-NPA2005},
	
	\begin{align}
		\begin{aligned}
			M_n^f(0^+)&\equiv\langle 0^+_f\|\sum_{j}\tau^+_j\sigma_j\| 1^+_n\rangle,\\ M_n^i(0^+)&\equiv\langle 1^+_n\|\sum_{k}\tau^+_k\sigma_k\| 0^+_i\rangle.
		\end{aligned}
	\end{align}
	
	The SSD phase-space factor (PSF) is given by
	
	\begin{align}
		\label{eq:PSFSSD}
		\begin{aligned}
			G^{2\nu}_{\rm{SSD}}&=\frac{(G_{F}\left|V_{ud}\right|)^4}{8\pi^7m_e^{2}\ln (2)}\int_{m_e}^{E_i-E_f-m_e}\int_{m_e}^{E_i-E_f-E_{e_1}}\\
			&\times \int_{0}^{E_i-E_f-E_{e_1}-E_{e_2}}p_{e_1}E_{e_1}p_{e_2}E_{e_2}\\
			&\times F_{ss}(E_{e_1})F_{ss}(E_{e_2})\\
			&\times \left[1+\eta^T(E_{e_1})\right]R(E_{e_1},E_i-E_f-m_e)\\
			&\times \left[1+\eta^T(E_{e_2})\right]  R(E_{e_2},E_i-E_f-E_{e_1}) \\
			&\times E_{\nu_1}^2E_{\nu_2}^2\frac{K^2+L^2+KL}{3}dE_{\nu_1}dE_{e_2}dE_{e_1}.
		\end{aligned}
	\end{align}
	It's important to note that in the above definition, we have again included the exchange and radiative corrections, which are discussed in details in Section~\ref{sec:ExRa}. The dimensionless quantities $K$ and $L$ are given by,
	
	\begin{align}
		\begin{aligned}
			K&=\frac{m_e\left[E_1 - (E_i+E_f)/2\right]}{\left[E_1 - (E_i+E_f)/2\right]^2-\varepsilon^2_{K}},\\ 
			L&=\frac{m_e\left[E_1 - (E_i+E_f)/2\right]}{\left[E_1 - (E_i+E_f)/2\right]^2-\varepsilon^2_{L}},
		\end{aligned}
	\end{align}
	where 
	
	\begin{align}
		\begin{aligned}
			\varepsilon_{K}&=\left(E_{e_2}+E_{\nu_2}-E_{e_1}-E_{\nu_1}\right)/2, \\
			\varepsilon_{L}&=\left(E_{e_1}+E_{\nu_2}-E_{e_2}-E_{\nu_1}\right)/2.
		\end{aligned}
	\end{align}
	
	An advantage of the SSD hypothesis is that $M^{2\nu}_{GT}(1)$ can be directly connected with the electron capture (EC) and the $\beta^-$-decay of the first $1^+$ state of the intermediate nucleus. In the case of $2\nu\beta\beta$-decay of $^{100}$Mo, this state is actually the ground state of the intermediate nucleus, $^{100}$Tc. Taking into account the recent experimental half-life for $2\nu\beta\beta$ decay of $^{100}$Mo, $T_{1/2}^{2\nu\rm{-expt}}=\left(7.07\pm0.11\right)\times 10^{18}~\rm{yr}$ \cite{Augier-PRL2023-100Mo}, and our calculation for the PSF, $G^{2\nu}_{\rm{SSD}}=4.008\times10^{-19}~ \rm{yr^{-1}}$, one can obtain,
	
	\begin{equation}
		\left(g^{\rm eff}_A\right)^2\left|M^{2\nu}_{GT}(1)\right|=\frac{1}{\sqrt{T_{1/2}^{2\nu\rm{-exp}}G^{2\nu}_{\rm{SSD}}}}=0.594\pm0.005
	\end{equation}
	On the other hand, $M_1^f(0^+)$ and $M_1^i(0^+)$ can be deduced from the $\log(ft)$ values of the EC and $\beta^-$-decay of the ground state of $^{100}$Tc, 
	
	\begin{equation}
		M_1^f(0^+)=\frac{1}{g^{\rm eff}_A}\sqrt{\frac{3D}{(ft)_{\beta^-}}},~~~M_1^i(0^+)=\frac{1}{g^{\rm eff}_A}\sqrt{\frac{3D}{(ft)_{\rm{EC}}}}
	\end{equation}
	where $D=\left[2\pi^3\ln(2)\right]/\left[(G_{F}\left|V_{ud}\right|)^2m_e^5\right]$. Taking into account that $\log (ft)_{\rm EC}=4.3\pm0.1$ and $\log (ft)_{\beta^-}=4.598\pm0.004$ \cite{Singh-NDS2021}, one can obtain,
	
	\begin{equation}
		\left(g^{\rm eff}_A\right)^2\left|M^{2\nu}_{GT}(1)\right|=\frac{3D}{\sqrt{(ft)_{\beta^-}(ft)_{\rm EC}}}=0.671^{+0.085}_{-0.076},
	\end{equation} 
	where the large variation interval primarily stems from the significant uncertainty in $\log (ft)_{\rm EC}$, for which a more precise measurement is desirable. The disagreement between the experimental NME and the one obtained from the $\log(ft)$ values, indicates that SSD hypothesis is not the whole story behind the $2\nu\beta\beta$ transition of $^{100}$Mo. Other higher $1^+$ states of $^{100}$Tc should contribute to the NME calculation. The discrepancy can potentially be elucidated by the cancellation effect arising from contributions of both low-lying and higher-lying states.
	
	Under the SSD hypothesis, the expressions for $\xi_{31}$ and $\xi_{51}$ are simplified, as only one term contributes in the summations in the NMEs, i.e.,
	
	\begin{align}
		\begin{aligned}
			&\xi_{31}^{\rm{SSD}}=\frac{4~ m_e^2}{\left[E_1 - (E_i+E_f)/2\right]^2},\\
			&\xi_{51}^{\rm{SSD}}=\frac{16~ m_e^4}{\left[E_1 - (E_i+E_f)/2\right]^4},
		\end{aligned}
	\end{align}
	In the particular case of the $2\nu\beta\beta$-decay of $^{100}$Mo, using the nuclear states energy differences $E_1-(E_i+E_f)/2=(E_1-E_i)+(E_i-E_f)/2=1.685$ MeV \cite{Singh-NDS2021}, one can find $\xi_{31}^{\rm{SSD}}=0.368$ and $\xi_{51}^{\rm{SSD}}=0.135$.

	Although the SSD hypothesis is just an assumption for the $2\nu\beta\beta$-decay, from the theoretical point of view, it has the advantage that the separation of the inverse half-life from Eq.~(\ref{eq:Half-lifeSSD}) is done without neglecting any leptonic energy dependence in the decay rate. Thus, the SSD observables can be used for verification of the ones obtained with the Taylor expansion formalism with fixed $\xi_{31}^{\rm{SSD}}=0.368$ and $\xi_{51}^{\rm{SSD}}=0.135$. The results are presented in Fig.~\ref{fig:ExactSSDComparison}, for the single (left panel) and summed (right panel) electron spectra for $2\nu\beta\beta$ decay of $^{100}$Mo. The results labeled with "exact SSD" are obtained from Eq.~(\ref{eq:PSFSSD}), and for the ones from Taylor expansion we assumed two scenarios: (i) up to next-to-next to leading order contributions enter in the total decay rate, i.e., $\Gamma^{2\nu} = \Gamma_0^{2\nu} + \Gamma_2^{2\nu}  + \Gamma_{22}^{2\nu} + \Gamma_4^{2\nu}$ (dashed curves), (ii) up to next to leading order contributions enter in the total decay rate, i.e., $\Gamma^{2\nu} = \Gamma_0^{2\nu} + \Gamma_2^{2\nu}$ (dotted curves). All spectra are normalized to unity, include radiative and exchange corrections, and are integrated with DHFS electron wave functions. As expected, including more terms in the Taylor expansion leads to a better agreement with the exact SSD for both single and summed electron spectra. This can be seem form the residuals depicted in the bottom panels of Fig.~\ref{fig:ExactSSDComparison}. A similar verification of the Taylor expansion formalism was conducted in our prior publication \cite{Nitescu-U2021} for the single-electron spectra of the $2\nu\beta\beta$-decay of $^{82}$Se, $^{100}$Mo, and $^{150}$Nd. However, it is worth noting that the additional corrections applied in this study were not considered in that previous work. Overall we observe a good agreement between the spectra, and we can consider that including up to next-to-next to leading order terms in the decay rate is quite precise for the description of the $2\nu\beta\beta$-decay of $^{100}$Mo. For higher experimental statistics, higher orders can be included.

	\subsection{The HSD hypothesis}
	
	In contrast to the SSD hypothesis, an alternative assumption is that the decay rate of $2\nu\beta\beta$-decay is primarily governed by the contributions from higher-lying states of the intermediate nucleus. This assumption is called the higher-state dominance (HSD) hypothesis. Given that the transition is mainly controlled by the GT resonance (GTR) states, typically situated at around $10-12$ MeV above the ground state of the initial nucleus, one can safely separate the nuclear part and the integration over the phase-space,
	
	\begin{equation}
		\label{eq:Half-lifeHSD}
		\left[T_{1/2}^{2\nu}\right]^{-1}=\left(g^{\rm eff}_A\right)^4\left|M^{2\nu}_{GT}\right|^2 G^{2\nu}_{0}.
	\end{equation} 
	Because no experimental information is available for the EC and $\beta^-$ decay of the GTR state of the intermediate nucleus, one can approximate,
	
	\begin{align}
		\begin{aligned}
			\left(g^{\rm eff}_A\right)^2\left|M^{2\nu}_{GT}\right|=\frac{1}{\sqrt{T_{1/2}^{2\nu\rm{-exp}}G^{2\nu}_{0}}}=0.202\pm0.002,
		\end{aligned}
	\end{align}
	where the $G^{2\nu}_{0}$ calculated in this work (see Table~\ref{tab:PSFDifferentWF}) is the leading order PSF from the Taylor expansion formalism.

	Assuming that only one state from the GTR region, with energy $E_{\rm{GTR}}$, dominates the transition, the ratios of the NMEs from the Taylor expansion formalism can be estimated from,
	
	\begin{align}
		\begin{aligned}
			&\xi_{31}^{\rm{HSD}}=\frac{4~ m_e^2}{\left[E_{\rm{GTR}} - (E_i+E_f)/2\right]^2},\\
			&\xi_{51}^{\rm{HSD}}=\frac{16~ m_e^4}{\left[E_{\rm{GTR}} - (E_i+E_f)/2\right]^4},
		\end{aligned}
	\end{align}
	For $E_{\rm GTR}>10$ MeV and consequently $(E_{\rm GTR}-E_i)+(E_i-E_f)/2>11.685$ MeV, it is evident that the ratios are negligible. Specifically, for the $2\nu\beta\beta$-decay of $^{100}$Mo, we have $\xi_{31}^{\rm{HSD}}<7.6\times10^{-3}$ and $\xi_{51}^{\rm{SSD}}<5.9\times10^{-5}$. We note that a more precise energy for the GTR can be obtained from \cite{Suhonen-NPA1988}, but the conclusion above remains true. It is also worth noting that a toy model was proposed in \cite{Nitescu-AIPCP2024} to describe the $2\nu\beta\beta$-decay NME as a mixture of contributions from the first state and one GTR state of the intermediate nucleus.

	\section{Exchange and radiative corrections in $2\nu\beta\beta$-decay}
	\label{sec:ExRa}
	
	We account for the possibility that one electron produced in the $2\nu\beta\beta$-decay is created in a bound orbital of the final atom, which corresponds to an occupied orbital in the initial atom. Simultaneously, an atomic electron from that bound orbital undergoes a transition to a continuum state of the final atom. This electron exchange, known as the atomic exchange correction, has been extensively studied in $\beta$-decay \cite{Bahcall-PR1963,Haxton-PRL1985,Harston-PRA1992,Pyper-PRSLA1998,Mougeot-PRA2012,Mougeot-PRA2014,Hayen-RMP2018,Hayen-arxiv2020,Haselschwardt-PRC2020,Nitescu-PRC2023}. In what follows, we extend the investigation of the exchange correction to the $2\nu\beta\beta$-decay, considering the exchange effect for both emitted electrons. Additionally, we incorporate the first-order radiative correction arising from the exchange of a virtual photon or the emission of a real photon during the $2\nu\beta\beta$-decay.
	
	The leading order radiative correction is given by \cite{Hayen-RMP2018},
	
	\begin{equation}
		R(E_e,E_e^{\textrm{max}})=1+\frac{\alpha}{2\pi}g(E_e,E_e^{\textrm{max}})
	\end{equation}
	where the function $g(E_e,E_e^{\textrm{max}})$ is given by \cite{Sirlin-PR1967,Sirlin-RMP2013}
	
	\begin{widetext}
	\begin{align}
		\begin{aligned}
			g(E_e,E_e^{\textrm{max}})&=3\ln(m_p)-\frac{3}{4}-\frac{4}{\beta}\textrm{Li}_2\left(\frac{2\beta}{1+\beta}\right)
			+\frac{\tanh^{-1}\beta}{\beta}\left[2\left(1+\beta^2\right)+\frac{\left(E_e^{\textrm{max}}-E_e\right)^2}{6E_e^2}-4\tanh^{-1}\beta\right]\\
			&+4\left(\frac{\tanh^{-1}\beta}{\beta}-1\right)\left[\frac{E_e^{\textrm{max}}-E_e}{3E_e}-\frac{3}{2}+\ln\left[2\left(E_e^{\textrm{max}}-E_e\right)\right]\right],
		\end{aligned}
	\end{align}
	\end{widetext}
	Here, $\beta=p_e/E_e$, $E_e^{\textrm{max}}$ is the maximum total energy of the electron, $m_p$ is the proton mass and $\textrm{Li}_2(x)$ is the dilogarithm function.
	
	The atomic exchange correction is given by \cite{Harston-PRA1992,Pyper-PRSLA1998},
	
	\begin{eqnarray}
		\label{eq:TotalEchangeCorrection}
		\eta^T(E_e)&=&f_s(2T_{s}+T_{s}^2)+(1-f_s)(2T_{\bar{p}}+T_{\bar{p}}^2)\nonumber\\
		&=&\eta_{s}(E_e)+\eta_{\bar{p}}(E_e)
	\end{eqnarray} 
	Here,
	
	\begin{eqnarray}
		f_s=\frac{g'^2_{-1}(E_e,R)}{g'^2_{-1}(E_e,R)+f'^2_{+1}(E_e,R)},
	\end{eqnarray}
	where $R=1.2A^{1/3}$ is the nuclear surface in fm, and $g_{\kappa}(E_e,r)$ and $f_{\kappa}(E_e,r)$ are the large- and small-component radial wave functions, respectively, for electrons emitted in the continuum states. All primed quantities pertain to the final atomic system. The continuum states are uniquely identified by the relativistic quantum number $\kappa$ and the energy $E_e$ and the bound orbitals are uniquely identified by the principle quantum numbers $n$ and $\kappa$. The quantities $T_{s}$ and $T_{\bar{p}}$ depend respectively on the overlaps between the bound $s_{1/2}$ ($\kappa=-1$) and $\bar{p}\equiv p_{1/2}$ ($\kappa=1$) orbitals wave functions in the initial state atom and the continuum states wave functions in the final state atom,
	
	\begin{eqnarray}
		\label{eq:TnsQuantities}
		T_{s}=\sum_{(ns)'}T_{ns}=-\sum_{(ns)'}\frac{\langle\psi'_{E_es}|\psi_{ns}\rangle}{\langle\psi'_{ns}|\psi_{ns}\rangle}\frac{g'_{n,-1}(R)}{g'_{-1}(E_e,R)}
	\end{eqnarray}
	and
	
	\begin{eqnarray}
		\label{eq:TnpQuantities}
		T_{\bar{p}}=\sum_{(n\bar{p})'}T_{n\bar{p}}=-\sum_{(n\bar{p})'}\frac{\langle\psi'_{E_e\bar{p}}|\psi_{n\bar{p}}\rangle}{\langle\psi'_{n\bar{p}}|\psi_{n\bar{p}}\rangle}\frac{f'_{n,+1}(R)}{f'_{+1}(E_e,R)}.
	\end{eqnarray}
	where $g_{n,\kappa}(r)$ and $f_{n,\kappa}(r)$ are the large- and small-component radial wave functions, respectively, for bound electrons. The summations in $T_s$ and $T_{\bar{p}}$ are performed over all occupied orbitals of the final atom, which, under the sudden approximation, correspond to the electronic configuration of the initial atom.
	
	In the calculation of the exchange correction, it is crucial to maintain orthogonality between the continuum and bound wave functions of the electron in the final atomic system, i.e., $\langle\psi'_{E_e\kappa}|\psi'_{n\kappa}\rangle=0$ \cite{Harston-PRA1992,Pyper-PRSLA1998}. We have demonstrated that non-orthogonal states have a significant impact on the overall behavior of the exchange correction with respect to the kinetic energy of the emitted electron. Therefore, we are adopting the same procedure as described in \cite{Nitescu-PRC2023,Nitescu-PRC2024}, i.e., modified self-consistent DHFS framework, to obtain the electron wave functions. 
	
	We emphasize that the exchange correction in $2\nu\beta\beta$ decay differs from that in $\beta$ decay. This discrepancy arises due to the fact that the initial nucleus undergoes a charge change of two units in the $2\nu\beta\beta$ decay process, whereas in $\beta$ decay, the charge change is only by one unit. Therefore, despite using the same notation (with single primes for the final states) as in \cite{Nitescu-PRC2023}, the overlaps described in Eqs.~(\ref{eq:TnsQuantities}) and (\ref{eq:TnpQuantities}) involve the wave functions for atomic systems with charge change of two units. 
	
	\begin{figure}
		\centering
		\includegraphics[width=0.45\textwidth]{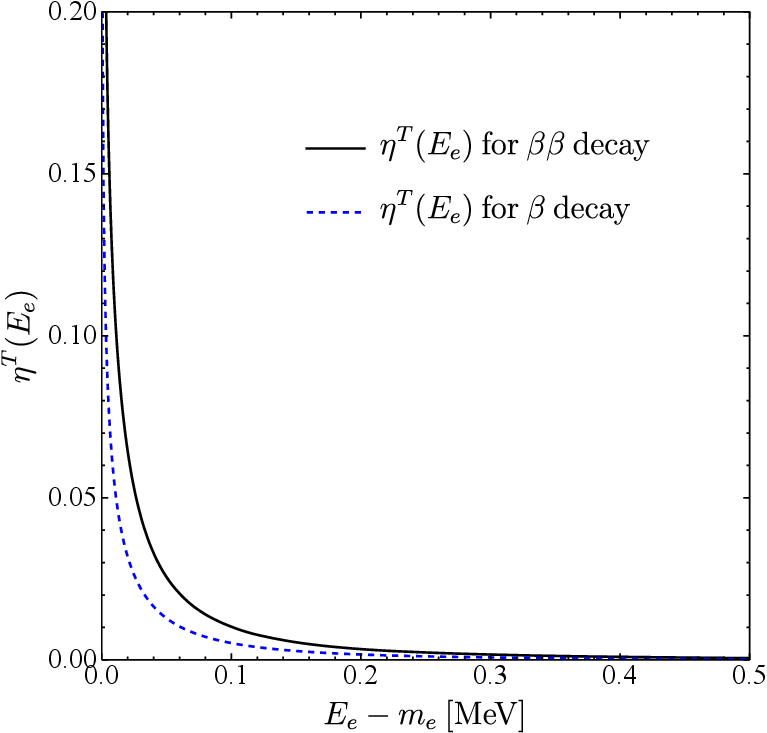}
		\caption{The total atomic exchange correction for $\beta$ decay (dashed blue) and $2\nu\beta\beta$ decay (solid black) of molybdenum. \label{fig:Exchange}}
	\end{figure}

	\begin{figure*}
		\centering
		\includegraphics[width=0.9\textwidth]{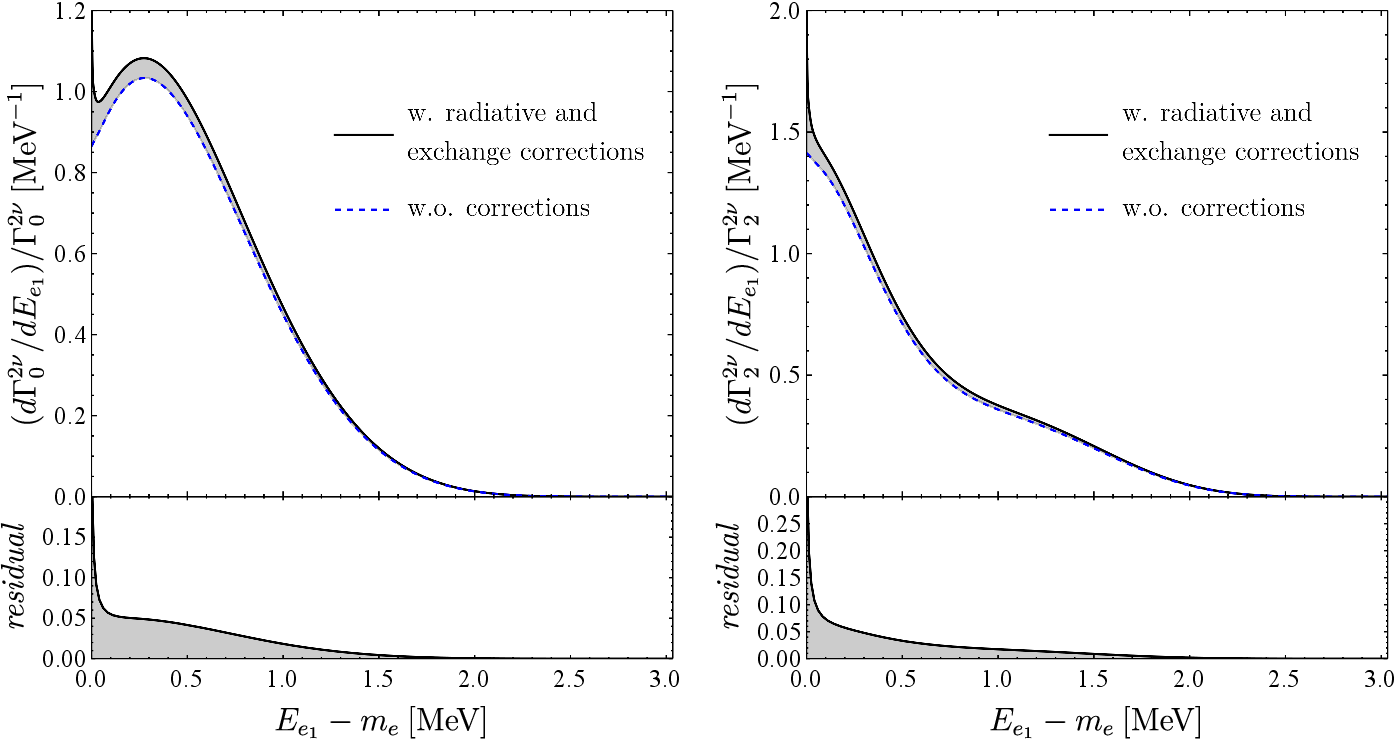}
		\caption{The first two contributions to the single electron spectrum of $2\nu\beta\beta$ decay of $^{100}$Mo, i.e., $d\Gamma_0/dE_e$ and $d\Gamma_2/dE_e$ with (solid curve) and without (dashed curve) radiative and exchange corrections. Only the uncorrected spectra are normalized to unity and the lower panels present the residuals.\label{fig:SingleBoth}}
	\end{figure*}
	
	In Fig.~\ref{fig:Exchange}, we depict the total exchange correction for one electron emitted in $2\nu\beta\beta$-decay (solid) and $\beta$-decay (dashed) of molybdenum isotopes. It can be observed that for $2\nu\beta\beta$-decay, the exchange effect is larger than that for $\beta$ decay across the entire energy range. The analytical parametrization introduced in \cite{Nitescu-PRC2023} was employed to obtain $\eta^{T}$ for $\beta$-decay. It is important to note that $^{100}$Mo does not undergo $\beta$-decay. However, the variation of the exchange correction with the mass number is relatively small. Therefore, the corrections presented in Fig.~\ref{fig:Exchange} can be applied to all isotopes of molybdenum.

	\section{The observables of the $2\nu\beta\beta$-decay of $^{100}$Mo including radiative and exchange corrections}

	We first examine the modifications in the shape of the single electron spectra in the $2\nu\beta\beta$ decay of $^{100}$Mo introduced by the radiative and exchange corrections. Specifically, our focus is on the first two contributions resulting from the Taylor expansion formalism, namely $d\Gamma_0/dE_e$ and $d\Gamma_2/dE_e$. The cumulative effect from both corrections is presented in Fig.~\ref{fig:SingleBoth}. The spectra without corrections (dashed) are normalized to unity, while the spectra with corrections (solid) are normalized to the decay rate of the uncorrected distribution. A pronounced modification appears in the low-energy region (0–100 keV), where the atomic exchange correction causes a steep rise. This result is consistent with previous observations in $\beta$-decay studies \cite{Harston-PRA1992,Pyper-PRSLA1998,Nitescu-PRC2023,Nitescu-PRC2024}. In contrast to the effect induced by the exchange correction, the shape of the spectra remains largely unaffected by the radiative correction, as indicated by the residuals (lower panels) closely following the distributions of the spectra. This suggests that while atomic exchange corrections have a substantial effect on the low-energy shape, radiative corrections mostly influence the overall decay rate.

	\begin{figure*}
		\centering
		\includegraphics[width=0.9\textwidth]{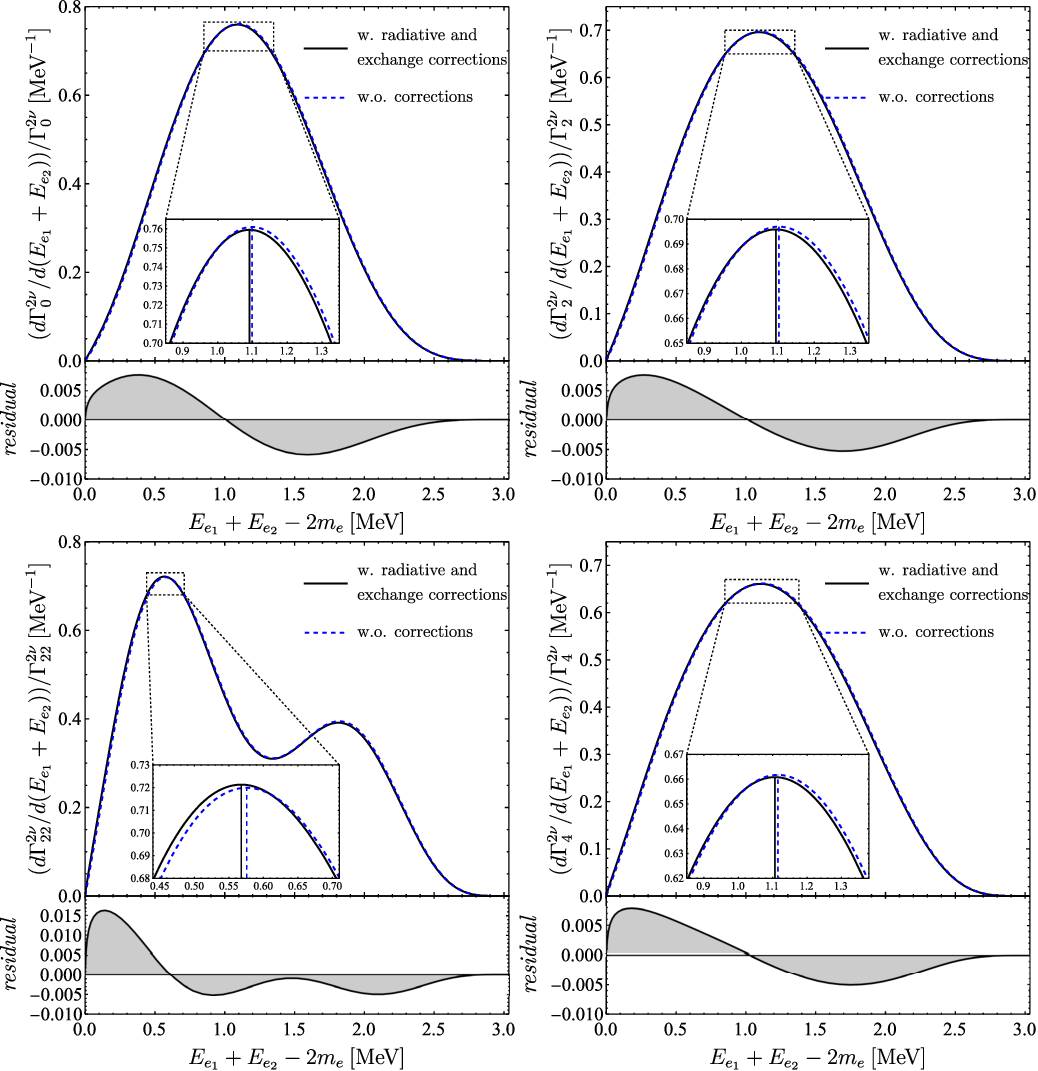}
		\caption{The uncorrected (dashed curve) and corrected (solid curve) summed electron spectra for $2\nu\beta\beta$ decay of $^{100}$Mo. In the corrected spectra both radiative and exchange corrections are included. All spectra are normalized to unity, and the lower panels display the residuals. The insets provide a closer view of the maxima of the spectra, with vertical lines intersecting at those points.  \label{fig:SummedBoth}}
	\end{figure*}

	\begin{table*}
		\caption{ \label{tab:PSFDifferentWF} The phase space factors $G^{2\nu}_{N}$ with $N={0,2,22,4}$ for the $2\nu\beta\beta$ decay of $^{100}$Mo. The results are obtained from Ref. \cite{Nitescu-U2021} (first row) and within the DHFS framework without additional corrections (second row). The third row corresponds to the results with the exchange correction, the fourth includes the radiative correction, and the fifth presents the calculations with both corrections. The last row displays the percentage deviations from the uncorrected DHFS results due to the radiative and exchange corrections. All values are given in units of $\textrm{yr}^{-1}$.} 
		\centering
		\begin{ruledtabular}
			\begin{tabular}{cccccc}
				Nucleus&\multicolumn{1}{c}{Correction(s)}&\multicolumn{1}{c}{$G_0^{2\nu}$}&\multicolumn{1}{c}{$G_2^{2\nu}$}&\multicolumn{1}{c}{$G_{22}^{2\nu}$}&\multicolumn{1}{c}{$G_4^{2\nu}$}\\
				\hline
				$^{100}$Mo	&\multicolumn{1}{c}{Ref. \cite{Nitescu-U2021}}&$3.279\times10^{-18}$&$1.498\times10^{-18}$&$1.972\times10^{-19}$&$8.576\times10^{-19}$\\
				&\multicolumn{1}{c}{DHFS}&$3.307\times10^{-18}$&$1.511\times10^{-18}$&$1.989\times10^{-19}$&$8.652\times10^{-19}$\\
				&\multicolumn{1}{c}{Exchange}&$3.343\times10^{-18}$&$1.536\times10^{-18}$&$2.031\times10^{-19}$&$8.835\times10^{-19}$\\
				&\multicolumn{1}{c}{Radiative}&$3.432\times10^{-18}$&$1.568\times10^{-18}$&$2.066\times10^{-19}$&$8.974\times10^{-19}$\\
				&\multicolumn{1}{c}{Radiative and Exchange}&$3.470\times10^{-18}$&$1.593\times10^{-18}$&$2.109\times10^{-19}$&$9.164\times10^{-19}$\\
				&\multicolumn{1}{c}{$\delta$}&$4.91\%$&$5.42\%$&$5.97\%$&$5.92\%$\\
			\end{tabular}
		\end{ruledtabular}
	\end{table*}

	\begin{figure*}
		\centering
		\includegraphics[width=0.9\textwidth]{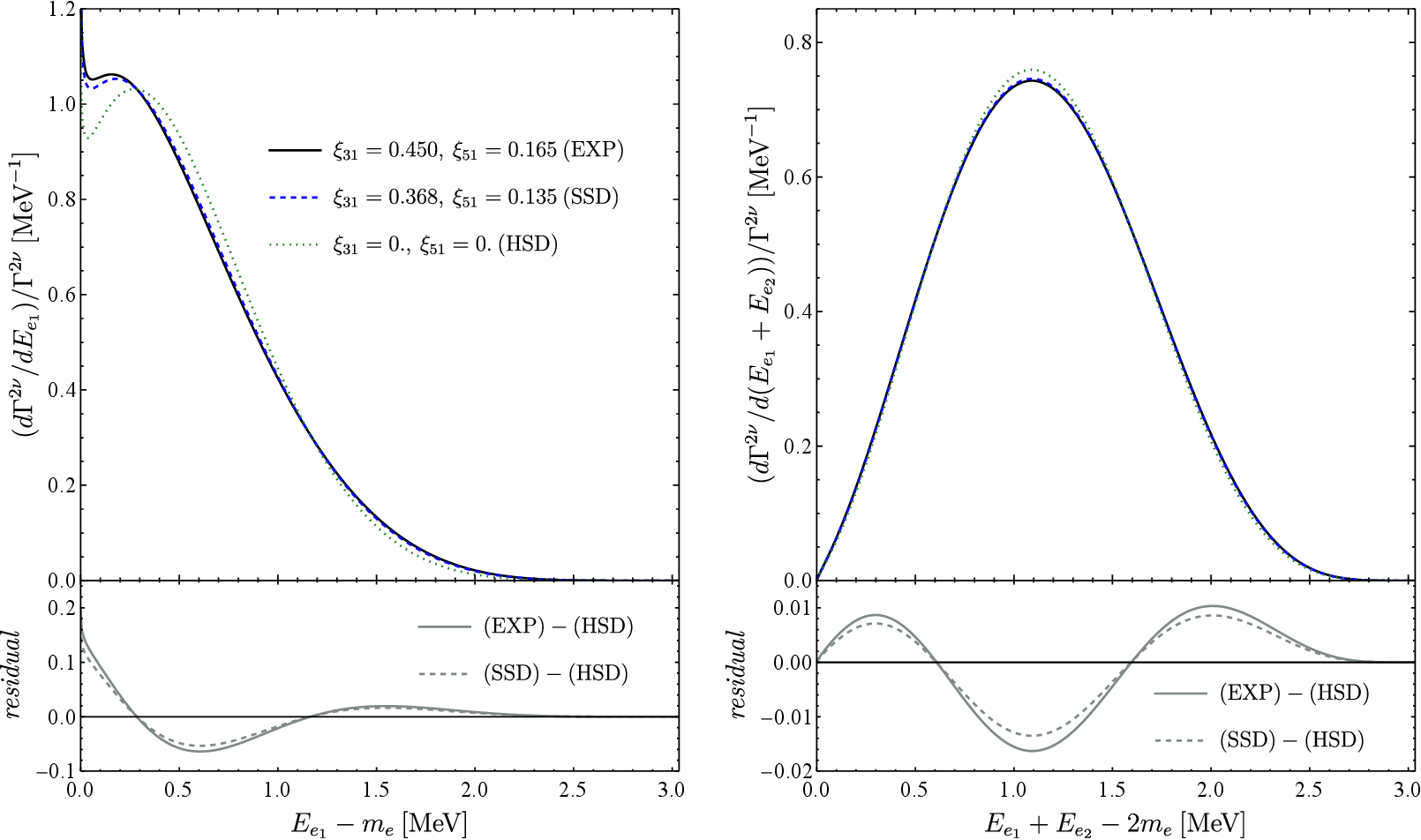}
		\caption{The corrected total single (left panel) and total summed (right panel) electron spectra for $2\nu\beta\beta$ decay of $^{100}$Mo, for different values of the ratios of NMEs: $\xi_{31}=0$ and $\xi_{51}=0$ which correspond to the HSD hypothesis (dotted line); $\xi_{31}=0.368$ and $\xi_{51}=0.135$ which correspond to the SSD hypothesis (dashed line); $\xi_{31}=0.450$ and $\xi_{51}=0.165$ from the experimental (EXP) measurement \cite{Augier-PRL2023-100Mo} (solid line). All spectra are normalized to unity, and the lower panels display the residuals between the spectra corresponding to the experimental and HSD values (solid line) and between the spectra corresponding to the SSD and HSD values (dashed line).  \label{fig:DifferentXi}}
	\end{figure*}

	The effect of radiative and exchange corrections on the total decay rate of $2\nu\beta\beta$-decay in $^{100}$Mo can be analyzed by reviewing the PSFs in Table~\ref{tab:PSFDifferentWF}. For comparison, we consider our previous calculations from \cite{Nitescu-U2021} (first row), where the atomic screening correction was incorporated as the solution of the Thomas-Fermi equation, as in \cite{Kotila-PRC2012,Stoica-PRC2013,Mirea-RRP2015,Stefanik-PRC2015,Simkovic-PRC2018,Stoica-FP2019}. The second row corresponds to the results obtained within the self-consistent DHFS framework. The deviations between the first and second rowa are solely attributed to a more precise treatment of the atomic screening correction in the DHFS method, providing a more accurate description of the atomic structure of the final system. The next three rows in Table~\ref{tab:PSFDifferentWF} show calculations that include the radiative correction, the exchange correction, and both corrections together. The final row provides the percentage change between the corrected and uncorrected PSFs using the DHFS framework. Based on our analysis, it can be concluded that the radiative and exchange corrections introduce modifications to each contribution in the decay rate of approximately 5\% for $2\nu\beta\beta$-decay of $^{100}$Mo. We note that the predominant correction driving the increase in the decay rate is the radiative correction.

	The influence of the radiative and exchange corrections on the shape of the summed energy electron spectra for $2\nu\beta\beta$-decay of $^{100}$Mo was investigated. The first four contributions to the summed electron spectrum are presented in Fig.~\ref{fig:SummedBoth}. It is observed that when both corrections are included in the calculation, all spectra exhibit a shift towards the left. The increase in events in the left half of the spectra is a combined effect of both corrections, which tend to shift the maxima in the same direction. Consequently, no cancellations between the corrections are obtained. In the insets from Fig.~\ref{fig:SummedBoth}, a closer examination of the shifts of the maxima of the spectra, indicated by the intersecting vertical lines, is presented. Although the shifts of the maxima are relatively small (approximately 10 keV), they become significant when considering the current and future experimental statistics. These effects may have implications for the experimental constraints on the strength parameters that govern various new physics scenarios, as the expected fingerprint of this scenarios is a shift in the maximum of the SM distribution.

	Finally, we investigate the impact of different sets of values for $\xi_{31}$ and $\xi_{51}$ on the total corrected single and summed electron spectra, which can be obtained from Eq.~(\ref{eq:TotalDecayRate}) taking into account that the PSFs are defined in Eq.~(\ref{eq:PhaseSpaceFactorsWithCorrections}). We consider three distinct assumptions: (i) the HSD hypothesis, in which $\xi^{\textrm{HSD}}_{31}=0$ and $\xi^{\textrm{HSD}}_{51}=0$, (ii) the SSD hypothesis, in which $\xi^{\textrm{SSD}}_{31}=0.368$ and $\xi^{\textrm{SSD}}_{51}=0.135$, and (iii) the experimental measurements of CUPID-Mo collaboration for $\xi^{\textrm{EXP}}_{31}=0.450$ and $\xi^{\textrm{EXP}}_{51}=0.165$ \cite{Augier-PRL2023-100Mo}. The outcomes are presented in the left panel of Fig.~\ref{fig:DifferentXi} for the single electron spectrum and in the right panel for the summed electron spectrum. All distributions are normalized to unity, incorporating both the radiative and exchange corrections. Notably, the single electron spectrum exhibits a high sensitivity to the values of $\xi_{31}$ and $\xi_{51}$ parameters. The differences between various assumptions become more pronounced in the low-energy region of the spectra, but unfortunately, these events are challenging to detect experimentally. This can be also seen from the bottom panel, which presents the residuals between different assumptions. For the summed electron spectrum, the influence of different $\xi_{31}$ and $\xi_{51}$ parameters is more pronounced in the maximum of the spectrum, at around $1.1$ MeV for $2\nu\beta\beta$-decay of $^{100}$Mo, as can also be seen from residuals in the bottom panel. Additionally, one can see that the variation of the summed ditribution with $\xi_{31}$ and $\xi_{51}$ values is also not negligible in the regions around $0.3$ MeV and $2$ MeV. These regions are experimentally accessible in all experiments, and the acquisition of these events do not require a tracking system for individual electrons, as in the case of the single electron spectra. 
	
	\vspace{-0.1cm}
	\section{Conclusions}

	In this paper, we increased the precision of predictions for observables in the $2\nu\beta\beta$-decay of $^{100}$Mo by incorporating radiative and atomic exchange corrections. As these corrections are introduced on top of our previous Taylor expansion formalism, we presented a connection between this approach and the SSD and HSD hypotheses. Additionally, we demonstrated that while the SSD hypothesis is an approximation for separating the decay rate, it remains useful in testing the truncation order of the Taylor series. To calculate the atomic exchange correction we employed a modified DHFS self-consistent framework that ensures orthogonality between continuum and bound states. We found that the exchange effect for one electron emitted in $\beta\beta$-decay is larger than in $\beta$-decay, as the atomic system's charge changes by two units in the former case.   
	  
	We found a steep increase in the number of event in the low-energy region of the single electron distribution due to the atomic exchange correction, which is in accordance with the previous studies on $\beta$-decay. Although the radiative correction leave the shape of the single electron spectrum unchanged, it is responsible for an overall increase in the decay rate of about 5\%. We also found that the both correction contribute constructively to a leftward shift of the maximum in the summed electron spectrum, amounting to about 10 keV for the $2\nu\beta\beta$-decay of $^{100}$Mo. Since similar shifts are predicted by new physics scenarios in $2\nu\beta\beta$-decay, our finding might influence the experimental constrains of the BSM parameters. Additionally, this corrections might affect the future $\xi_{31}$ and $\xi_{51}$ measurements. Finally, we provided the corrected single and summed electron spectra for the $2\nu\beta\beta$-decay of $^{100}$Mo under the assumptions of the SSD and HSD hypotheses, as well as for experimentally measured values of the $\xi_{31}$ and $\xi_{51}$ parameters.
	
	\section*{Acknowledgments} The authors thank Volodymyr Tretyak for valuable and fruitful discussion during the early stages of this work. O.N. acknowledges support from the Romanian Ministry of Research, Innovation, and Digitalization through Project No. PN 23 21 01 01/2023.  F. \v{S}. acknowledges support by the Slovak Research and Development Agency under Contract No. APVV-22-0413, VEGA Grant Agency of the Slovak Republic under Contract No. 1/0618/24 and the Ministry of Education, Youth and Sports of the Czechia under the INAFYM Grant No. CZ.02.1.01/0.0/0.0/16\_019/0000766.

	\bibliographystyle{apsrev4-2}
	\bibliography{bibliography}

\end{document}